%% file: triplets.tex
\documentclass[]{webofc}
\usepackage[varg]{txfonts}   % Web of Conferences font
\usepackage{amsmath,amssymb,graphicx,slashed}

\newcommand{\beq}{\begin{equation}}
\newcommand{\eeq}{\end{equation}}

\newcommand{\nn}{\nonumber \\}

\newcommand{\GSM}{$\mathrm{G_{SM}}$}
\newcommand{\Zp}{Z$^\prime$}
\newcommand{\Wp}{W$^\prime$}

\newcommand{\calL}{\mathcal{L}}
\newcommand{\calW}{{\mathcal{W}}}

\def\MET{\slashed{E}_T}

\newcommand{\tinymath}[1]{{\tiny{\mbox{$#1$}}}}

\woctitle{LHCP 2013}
\begin{document}

\renewcommand{\thefootnote}{\fnsymbol{footnote}}
\setcounter{footnote}{1}

\title{Vector triplets at the LHC\thanks{Talk given by M. P\'erez-Victoria at LHCP 2013, Barcelona, Spain, May 13-18, 2013.}}
\renewcommand{\thefootnote}{\arabic{footnote}}
\setcounter{footnote}{0}

\author{Javier M.\ Lizana\inst{1}\fnsep\thanks{\email{jlizan@ugr.es}} \and
        Manuel P\'erez-Victoria\inst{1}\fnsep\thanks{\email{mpv@ugr.es}}
        % etc.
}

\institute{CAFPE and Departamento de F\'{\i}sica Te\'orica y del Cosmos\\
Universidad de Granada, E-18071 Granada, Spain}

\abstract{%
Several popular extensions of the Standard Model predict extra vector fields that transform as triplets under the gauge group $\mathrm{SU(2)_L}$. These multiplets contain $Z'$ and $W'$ bosons, with masses and couplings related by gauge invariance. We review some model-independent results about these new vector bosons, with emphasis on di-lepton and lepton-plus-missing-energy signals at the LHC.
}
\maketitle
\section{Introduction}
\label{intro}
Extra vector bosons are a common feature of all theories beyond the Standard Model (SM) with an extended gauge group. 
%This is the case, for instance, of Grand Unified Theories, Little Higgs models and many models in extra dimensions. 
The possible extensions of the gauge group have a common feature: they contain the SM group $\mathrm{G_{SM}}=\mathrm{SU(3)_C\times SU(2)_L \times U(1)_Y}$ as a subgroup. It is convenient to use this piece of information to systematically organize the analyses of the new vectors from a model-independent point of view~\cite{delAguila:2010mx}. In this phenomenological approach, gauge invariance under \GSM\ plays two crucial, related roles: it provides a classification principle and it restricts the possible interactions of the new particles, gi\-ving rise to a simple and natural parameterization in terms of masses and couplings.

Assuming renormalizable interactions to avoid suppressions from a higher scale, gauge invariance implies that only fifteen different multiplets of vector bosons can be singly produced at colliders~\cite{delAguila:2010mx}. Each of these multiplets has definite quantum numbers under \GSM.\footnote{We use the standard notation $(C,I)_Y$ to denote irreducible representations under \GSM, with $C$ and $I$ the dimension of the $\mathrm{SU(3)_C}$ and $\mathrm{SU(2)_L}$ representation, respectively, and $Y$ the hypercharge.}  We can have, for instance, singlets ${\cal B} \in (1,1)_0$, such as the $Z'$ bosons associated with an extra $U(1)$ factor, color octets ${\cal G} \in (8,1)_0$, such as the Kaluza-Klein excitations of gluons, etc. Here, we will study another important type of vector bosons: isospin triplets ${\cal W} \in (1,3)_0$. These multiplets are formed by a neutral $Z'$ boson and a pair of $W'$ bosons, of charge $\pm 1$. 

Vector triplets are interesting for various reasons. From the theoretical side, they appear in well-known extensions of the SM, such as little Higgs, composite and extra dimensional models.\footnote{The simplest gauge extension of the SM that gives rise to vector triplets is $\mathrm{SU(3)_C \times SU(2)_1\times SU(2)_2 \times U(1)_Y}$, spontaneously broken to \GSM. See~\cite{Schmaltz:2010xr} for an analysis of a simple model based on this pattern of symmetry breaking.} Experimentally,  they are the only vector bosons, together with singlets ${\cal B}$, that can give rise to sizable resonant signals with leptonic final states at the LHC~\cite{deBlas:2012qp}.\footnote{We are assuming that there are no extra fermions lighter than the vector bosons.} Unlike singlets, they have charged components that can contribute to lepton-plus-missing-transverse-energy ($\ell+\MET$) events. They can also produce observable di-jet and di-boson signals. 

The model-independent study of the collider phenomenology of vector triplets was initiated in~\cite{deBlas:2012qp}. Here, we will describe the basic properties of these vector bosons and review some results about their leptonic signatures at the LHC. We will emphasize the possibility of combining the data from searches of $Z'$ and $W'$ bosons in this context.

%Their $W'$ components may be seen in lepton-plus-missing-transverse-energy ($\ell$+\MET) events at the LHC. Actually, in the absence of additional light fermions, they are the unique heavy vector fields that can give a sizable contribution to this signal~\cite{deBlas:2012qp}. The reason is that only the $W'$ bosons in this multiplet can couple to both quarks and leptons directly.\footnote{Other charged vector bosons, like a $W'_R$ in the $(1,1)_1$ representation, can adquire these couplings via their mixings with the SM $W$ bosons. These mixings, however, must be small, due to constraints from electroweak precision data~\cite{delAguila:2010mx} and perturbativity (since they arise from electroweak breaking). The effects of this small mixing could be observed in di-boson final states~\cite{Grojean:2011vu}.} Similarly, there are only two kinds of $Z'$ bosons with significant contributions to di-lepton ($\ell^+\ell^-$) final states: singlets ${\cal B}$ and the neutral components of triplets ${\cal W}$. 

\section{Effective description of vector triplets}
\label{sec-eff}

Let us consider an extension of the SM with a vector field $\calW$ that belongs to the $(1,3)_0$ representation of \GSM. We make no assumption about the origin of the new field. To analyse the phenomenology of this scenario in a model-independent manner, we consider a general Lagrangian with gauge-invariant operators of dimension 4, built with the SM fields and $\calW$~\cite{delAguila:2010mx,Grojean:2011vu}:
\beq
\calL=\calL_\mathrm{SM} + \calL_\calW^0 + \calL_\calW^\mathrm{int},
\eeq
with $\calL_\mathrm{SM}$ the SM Lagrangian,
\begin{align}
&\calL_\calW^0 = -\frac{1}{2} [D_\mu \calW_\nu]^a [D^\mu \calW^\nu]_a + \frac{1}{2} [D_\mu \calW_\nu]^a [D^\nu \calW^\mu]_a  \nn
&\hspace{1cm}+ \frac{\mu^2}{2} \calW_\mu^a \calW^\mu_a  \label{LagW0},
\end{align}
and
\begin{align}
\calL_\calW^\mathrm{int} =  & g_{2} \calW_\mu^a \calW^\mu_a \phi^\dagger \phi - 
g_l \calW_a^\mu \bar{l}_i \gamma_\mu \frac{\sigma^a}{2} l_i - g_q \calW_a^\mu \bar{q}_i \gamma_\mu \frac{\sigma^a}{2} q_i  \nn
&-\left(i g_\phi \calW_a^\mu \phi^\dagger \frac{\sigma^a}{2} D_\mu \phi   + \mathrm{h.c.}\right)
+ \frac{1}{2} g_W \epsilon_{abc} \calW^a_\mu \calW^b_\nu W_{\mu\nu}^c .
\label{LagW1}
\end{align}
The derivative $D_\mu$ is covariant with res\-pect to the SM gauge group; $l_i$ and $q_i$ denote the left-handed lepton and quark doublets of the $i$th family, respectively; and $W_{\mu\nu}^a$ is the field-strength tensor of the $\mathrm{SU(2)_L}$ gauge fields. We have only written the terms with up to two extra vector fields, which are the ones relevant for phenomenology, and have assumed, for simplicity, diagonal and universal couplings to the three families of fermions in the interaction basis. Also for simplicity, we will consider in the following a real coupling $g_\phi$.

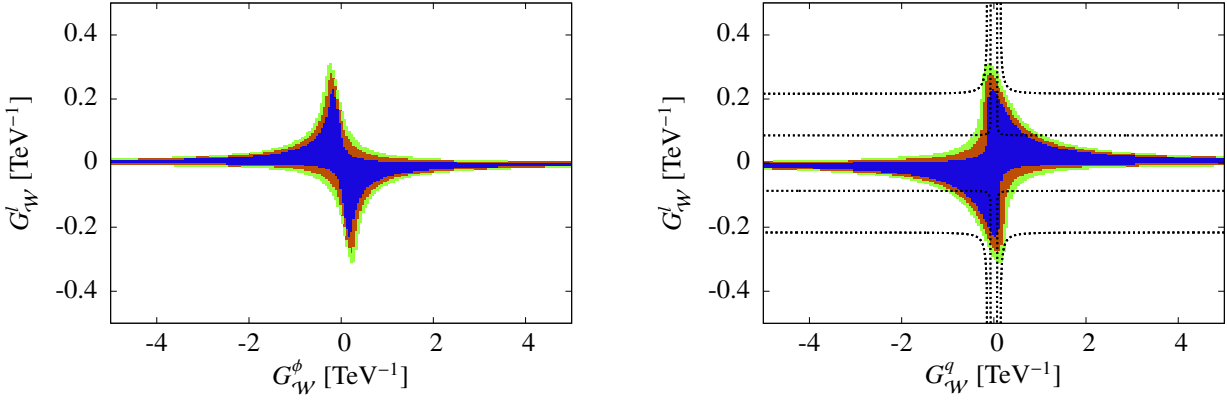
\begin{figure*}
\centering
\input{Tex_glgqgH_W}
\caption{Left: From darker to lighter, confidence regions with $\Delta \chi^2\leq2$ (blue), 4 (orange) and 6 ($95\%$ C.L.) (green), respectively, in the $G_{\cal W}^q=0$ plane. Right: The same in the $G_{\cal W}^\phi=0$ plane. We also plot the curves corresponding to constant values of $|\tilde{g}/\mu|=0.1$ (inner curve), 0.25 (outer curve) TeV$^{-1}$. 
}
\label{fig-EW}
\end{figure*}
Electroweak precision tests put limits on the parameters in this Lagrangian. They were calculated from a global fit in \cite{delAguila:2010mx}, and stay mostly unchanged when updates in the precision observables are taken into account. In Figure \ref{fig-EW}, we show the limits on the coupling to mass ratios $G^{l,q,\phi}_\calW \equiv g_{l,q,\phi}/\mu$ in two different planes. The flat direction along $G_{\calW}^\phi$, for vanishing $G_{\cal W}^l$, is due to the fact that the vector triplet preserves custodial symmetry and does not modify the $\rho$ parameter. However, note that in order to have leptonic events we need $G_{\cal W}^l \neq 0$. In this case, the plots show that the ratios $G^{\phi}_\calW$ and $G^{1}_\calW$ have upper bounds. In the second plot, we have also displayed lines of constant $|\tilde{g}/\mu|$, with 
\beq
\tilde{g} \equiv \frac{2 g_q g_l} {\sqrt{3g_q^2+g_l^2}}. 
\eeq
As we explain below, the LHC cross sections depend mostly on this combination of couplings. From the right-hand plot in Fig.~\ref{fig-EW}, we can read the upper bound $|{\tilde g}/\mu |\lesssim 0.25$ TeV$^{-1}$.

After electroweak symmetry breaking, the neutral and charged components of the triplet mix with the $Z$ and the $W$ gauge bosons. The resulting neutral mass matrix is
\beq
M^2_\mathrm{n} = \left( \begin{array}{cc}
M_{Z^0}^2 & \frac{x}{\cos \theta_W} \\
\frac{x}{\cos \theta_W} & M^2_\calW
\end{array} \right) ,
\eeq
with $M_{Z^0}^2=(g^2+g^{\prime 2}) v^2/4$, $\theta_W$ the Weinberg angle, $M^2_\calW = \mu^2 +g_{2} v^2$ and $x=g g_\phi v^2/4$. Up to terms of order $v^4/M_\calW^4$, the eigenvalues of this matrix are
\begin{align}
& M_Z^2 \simeq M_{Z^0}^2 - \frac{x^2}{M_\calW^2 \cos^2 \theta_W}, \nn
& M_{Z^\prime}^2 \simeq M^2_\calW + \frac{x^2}{M_\calW^2 \cos^2 \theta_W} ,
\end{align}
and the matrix is diagonalized by a rotation of angle $\alpha_n$ with $\sin \alpha_n \simeq \frac{g g_{\phi}}{4 \cos \theta_W} \frac{v^2}{M_\calW^2}$.
Similarly, the mass matrix in the charged sector is 
\beq
M^2_\mathrm{c} = \left( \begin{array}{cc}
M_{W^0}^2 & x \\
x & M^2_\calW
\end{array} \right) ,
\eeq
with $M_{W^0}^2=g^2 v^2/4$ the SM $W$ mass. Neglecting again terms of order $v^4/M_\calW^4$, the eigenvalues read
\begin{align}
& M_W^2 \simeq M_{W^0}^2 - \frac{x^2}{M_\calW^2}, \nn
& M_{W^\prime}^2 \simeq M^2_\calW + \frac{x^2}{M_\calW^2}, 
\end{align}
and the mixing angle in the $2\times2$ unitary matrix diagonalizing $M_c^2$ is $\sin \alpha_c \simeq \frac{g g^{\phi}_\calW}{4} \frac{v^2}{M_\calW^2}$. It is apparent that, as announced before, the mixing with the triplet does not spoil the custodial-symmetry relation between the mass of the Z and the W. Moreover, the neutral and charged heavy vectors stay nearly degenerate:
\beq
M_{Z^\prime} \simeq M_{W^\prime}+\frac{g^{\prime 2} g_\phi^2 \, v^4}{32M_{W^\prime}^3} .
\eeq
On the other hand, the mixing does modify the interactions of the mass eigenstates with the fermions, including the appearance of a new coupling of the $Z'$ to right-handed singlets. It also induces interactions involving one heavy vector eigenstate and light vectors or the Higgs boson. All these effects are suppressed by $\sin \alpha_{n,c}$, which in turn are constrained by perturbativity of the coupling $g_\phi$ and by the electroweak limits above. Note, however, that the decay widths of the heavy vectors into longitudinal light vector bosons and Higgs bosons are enhanced by derivative couplings, which can lead to a large di-boson branching ratio and an increased total width, even for relatively small mixing.

In order to simplify the LHC analysis, we will set $g_\phi$ to zero in the following, so that the mixing vanishes. This assumption maximizes the leptonic branching ratios. The phenomenology of our effective theory is then characterized by three parameters: $M_\calW$, $g_l$ and $g_q$. (The $g_W$ term plays no relevant role when $g_\phi=0$.)  The masses of the new vector bosons are degenerate, $M_{Z^\prime}=M_{W^\prime} =M_\calW$, \footnote{Within the approximation used to derive the electroweak bounds above, we can safely interchange $\mu$ and $M_\calW$.} while their couplings to fermions in the mass eigenstate basis are given by
\begin{align}
& \calL^\mathrm{CC}_\calW =  -\frac{1}{\sqrt{2}} \left(g_q \bar{u}_{Li}\gamma_\mu V_{ij} d_{Lj} + g_l  \bar{e}_{Li}\gamma_\mu  \nu_{Li} \right)  W^{\prime +}_\mu + 
\mathrm{h.c.}  ,  \label{chargedcurrent}  \\
& \calL^\mathrm{NC}_\calW =  -\frac{1}{2} \left[ g_q \left( \bar{u}_{Li}\gamma_\mu u_{Li} - \bar{d}_{Li} \gamma_\mu  d_{Li}  \right) \right. \nn
&\hspace{1cm}\left. + g_l \left(\bar{\nu}_{Li} \gamma_\mu \nu_{Li} - \bar{e}_{Li} \gamma_\mu e_{Li} \right) \right]  Z^{\prime}_\mu , \label{tripletZcouplings}
\end{align}
with $V$ the CKM matrix.

We see that the $W'$ and $Z'$ bosons only interact with the left-handed fermion chiralities. For this reason, the notation $W_L'$ is sometimes used to distinguish the charged vector in this multiplet from a $W_R'$, which couples instead to the right-handed fermions.\footnote{A $W_R'$ vector boson, which appears for instance in Left-Right models, belongs to the complex singlet ${\cal B}^1 \in (1,1)_1$. See \cite{delAguila:2010mx,Langacker:1989xa} for bounds on its parameters and \cite{Grojean:2011vu} for a model-independent analysis of its collider signatures. Although it does not couple to the SM neutrinos, this $W_R'$ could give $\ell+\MET$ signals in the presence of very light right-handed neutrinos.} Analogously, the $Z'$ in this multiplet may be called a $Z_L'$.

For the particular choice $g_l=g_q=g$, in the charged sector we have exactly the sequential $W'$ model, commonly used as a benchmark in the Atlas and CMS analyses of charged vector bosons. On the other hand, the $Z'$ is not sequential. It couples to the third component of isospin. The isospin dependent couplings of this neutral vector reveal that it belongs to a multiplet of dimension higher than one.

\section{Vector triplets at the LHC: leptonic signals}
\label{sec-eff}

For $g_\phi=0$, Drell-Yan is the only re\-le\-vant production mechanism of the $Z'$ and $W'$ bosons in the triplet. These heavy bosons can then decay into dif\-fe\-rent final states involving quarks or leptons. Here, we concentrate on the leptonic modes, $\ell^+ \ell^-$ and $\ell+\MET$ (with $\ell=e,\mu$). In the narrow width approximation, the corresponding cross sections at the LHC can be written as
\begin{align}
& \sigma(pp \to Z' \to \ell^+ \ell^-) = \frac{\pi}{6s} \left[ c_u \omega_u\left(s,m_{Z'}^2\right) + c_d \omega_d\left(s,m_{Z'}^2\right) \right] , \\
& \sigma(pp \to W' \to \ell^\pm \nu)=\frac{\pi}{6s} c_c \omega_c \left(s,m_{W'}^2 \right),
\end{align}
where the functions $\omega_{u,d,c}$, which contain the information of parton distribution functions, depend on the collider ener\-gy and the mass of the heavy bosons, but not on their couplings. This model-independent parameterization of the $Z'$ cross section was proposed in \cite{Carena:2004xs} (see also \cite{Accomando:2010fz} for an update for the LHC); the one for the $W'$ one is a trivial extension. The phenomenological parameters $c_{u,d,c}$ carry the information about the $Z'$ and $W'$ couplings. For vector triplets, they are given by~\cite{deBlas:2012qp}
\begin{align}
&c_{u} =c_d= \frac{\tilde{g}^2}{96} , \label{cud} \\
&c_c = \frac{\tilde{g}^2}{24} .  \label{cc}
\end{align}
Therefore, in the narrow width approximation, the cross sections for the leptonic processes mediated by the $Z'$ and $W'$ bosons in a vector triplet depend on the same two parameters: the effective coupling $\tilde{g}$, defined above, and the mass $M_{\cal W}=M_{Z'}=M_{W'}$.  There is thus a complete correlation between the $\ell^+\ell^-$ and $\ell+\MET$ events produced by the triplets. This simple property would be crucial to identify this SM extension in case of observation at the LHC.
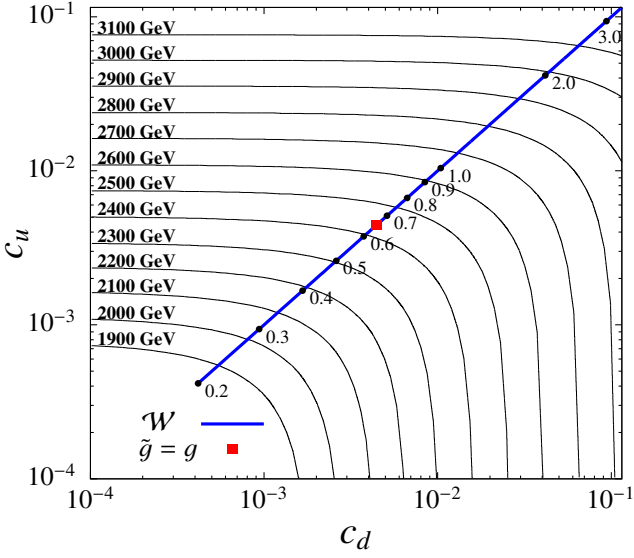
\begin{figure}
\centering
\input{cucd_W0}
%\vspace*{5cm}
\caption{$95\%$ C.L. exclusion limits for the cross section of the $Z'$ plotted over the $c_u-c_d$ plane, from~\cite{Chatrchyan:2012it}. The $c_u$ and $c_d$ values of the triplet as a function of $\tilde{g}$ are represented by the blue line.}
\label{fig_cucd}
\end{figure}
For the moment, we have to content ourselves with using the LHC results to put limits on the couplings and masses. In Fig.~\ref{fig_cucd}, we show the implications of the general bounds on $c_u$ and $c_d$ obtained by CMS in~\cite{Chatrchyan:2012it} from a di-lepton analysis, for the effective coupling $\tilde{g}$ of the coupling.

While the narrow width approximation captures well the basic implications of extra vector triplets, the cross sections can be significantly affected by the shape of the resonances and the interference with SM amplitudes. These effects are specially important for the $W'$, as emphasized in \cite{Accomando:2011eu}, and also for broad $Z'$ bosons with masses close to the kinematical reach of the collider \cite{Accomando:2013sfa}. The precise cross sections are thus sensitive to the individual couplings $g_l$ and $g_q$, and not only to the combination $\tilde{g}$. For instance, when $g_l$ and $g_q$ have the same sign, the interference of the amplitudes mediated by $W$ and $W'$ bosons is negative in the energy region between their poles, whereas for opposite signs, it is positive.
\begin{figure}
\centering
 \input{limtrp_gH0}
\caption{95$\%$ C.L. exclusion limits for the triplet model with $g_q=g_l=\tilde{g}$ and $g_\phi=0$. Regions above the curves are excluded. The bounds obtained using the $\ell^+\ell^-$ and $\ell+\slashed{E}_T$ channels separately are delimited by the (red) dashed and (blue) dot-dashed lines, respectively. The solid black line represents the limits for the combination of both channels. Finally, the grey bands represent the systematic uncertainty, corresponding to a $\pm10\%$ variation in the signal. We also show in blue the region excluded by electroweak precision data.}
\label{limitestr}       % Give a unique label
\end{figure}
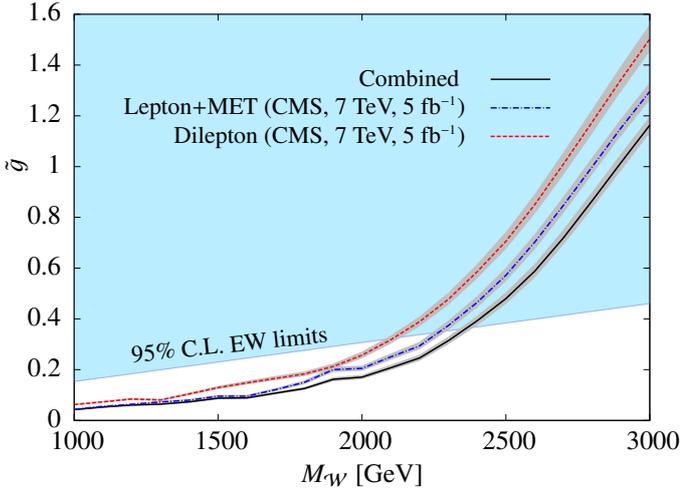

A complete general analysis can then be done (for $g_\phi=0$) in terms of three parameters (a mass and two couplings). Here, we just present precise bounds in the cases with $g_q=g_l$, for which we also have $\tilde{g}=g_l$.  We use data from direct searches of resonances decaying to leptons in CMS at the LHC (~\cite{Chatrchyan:2012it} and~\cite{Chatrchyan:2012meb}). We find limits from both the  $\ell^+ \ell^-$ and the $\ell+\slashed{E}_T$ channels. Moreover, in order to take advantage of the theoretical correlations between the \Zp{} and \Wp{} signals, we also find limits from a combination of the data in both channels, which involves a common test-statistic (for more details, see ~\cite{deBlas:2012qp}). Figure \ref{limitestr} shows our exclusion limits at the $95\%$ C.L., in the $\mathrm{CL_s}$ approach, for the individual $\ell^+ \ell^-$ and $\ell+\slashed{E}_T$ channels and for their combination. Note that each curve represents limits on the masses and couplings of both the $Z'$ and the $W'$ bosons in the triplet. For comparison, we also display in the same plot the region excluded by electroweak precision tests.

We can notice in this figure several interesting features.\footnote{We have used the results of the analysis in~\cite{deBlas:2012qp}, which were obtained with data from LHC at 7~TeV. However we expect all the qualitative features described in this paragraph to hold for the full LHC-7 and LHC-8 dataset.} First, comparing the limits from individual channels, we see that the ones from $\ell+\slashed{E}_T$ are stronger, despite the fact that this final state requires a transverse-mass, rather than invariant-mass, analysis. The reason is the factor-$4$ difference between $c_u=c_d$ and $c_c$, which comes from the couplings in Eqs.~(\ref{chargedcurrent}) and (\ref{tripletZcouplings}). This final state thus gives stronger constraints on the $Z'$ inside the triplet than the ones from the more obvious di-lepton final state. Second, we see in the plot that the combination of neutral and charged channels leads to stronger bounds than the ones from the best of the individual channels. In particular, this shows that, even if the combined limits are dominated by  $\ell+\slashed{E}_T$, the information from $\ell^+ \ell^-$ data is useful too. Finally, we see that in the region of large masses and couplings, the bounds from electroweak precision data are stronger than the ones from the searches of leptonic resonances at the LHC. For example, in the case of the sequential $W'$ model, the LHC has only explored so far masses that had already been excluded by LEP. So, it is no wonder that such a particle has not been discovered in the first LHC run.

%%%%%%%%%%%

\section{Conclusions}

Vector triplets appear in many models beyond the SM. We have showed how their phenomenology can be studied in a model-independent fashion, using an effective Lagrangian that describes their general interactions with the SM particles. The most characteristic signature of a vector triplet at the LHC is a pair of peaks at the same position (and of comparable size) in the invariant-mass and transverse-mass distributions of  $\ell^+ \ell^-$ and $\ell+\slashed{E}_T$ events, respectively. The combined analysis of these distributions is useful for limit setting, discovery and model identification.

Searches with other final states are important as well. A vector triplet with small lepton couplings is better seen as a resonance in di-jet events. This channel is sensitive to the quark couplings, which are poorly constrained by electroweak precision data. In this case, however, the $Z'$ and $W'$ bosons contribute inclusively to the same observable, so this channel alone would not allow to distinguish a vector triplet from many other possibilities. The $tt$ and $tb$ final states, on the other hand, receive separate contributions from the $Z'$ and $W'$ in the triplet, and would be useful to identify this multiplet. These decay modes are especially relevant when the triplets couple preferentially to the third family of quarks. Finally, the different di-boson final states become essential for non-negligible mixing. A combined analysis of neutral and charged channels is also possible in this case.

We have only considered here a minimal extension of the SM with one vector triplet. In general, all the results above are unchanged by the presence of additional multiplets of vector bosons. We should mention, however, one exception. In extensions with a singlet $\cal B$ and a triplet $\cal W$, a mixing between the $Z'$ bosons in the two multiplets is allowed in the electroweak broken phase~\cite{deBlas:2012qp}. This mixing has two effects: it removes the mass degeneracy of the triplet and it modifies the couplings of the eigenvectors. However, the mixing can only be sizable when the gauge-invariant masses of both multiplets are similar. As a consequence, one of the neutral eigenstates always stays close to the $W'$. This scenario can be used to construct a consistent sequential $Z'$ model. Such a consistent, gauge-invariant model necessarily predicts, besides the $Z'$ with SM couplings, another $Z'$ boson and a  $W'$ boson, all of them with a comparable mass. More details about extensions with vector triplets and singlets can be found in~\cite{deBlas:2012qp}. 

\section*{Acknowledgements}
MPV thanks the organizers of LHCP 2013 for a pleasant and interesting conference, and both authors thank Jorge de Blas for his collaboration in the study of vector triplets. This work has been supported by the MICINN projects FPA2006-05294 and FPA2010-17915, and by the Junta de Andaluc\'{\i}a projects FQM-101 and FQM-06552.

\end{document}

%% file: Tex_glgqgH_W.tex
% GNUPLOT: LaTeX picture with Postscript
\begingroup
  \makeatletter
  \providecommand\color[2][]{%
    \GenericError{(gnuplot) \space\space\space\@spaces}{%
      Package color not loaded in conjunction with
      terminal option `colourtext'%
    }{See the gnuplot documentation for explanation.%
    }{Either use 'blacktext' in gnuplot or load the package
      color.sty in LaTeX.}%
    \renewcommand\color[2][]{}%
  }%
  \providecommand\includegraphics[2][]{%
    \GenericError{(gnuplot) \space\space\space\@spaces}{%
      Package graphicx or graphics not loaded%
    }{See the gnuplot documentation for explanation.%
    }{The gnuplot epslatex terminal needs graphicx.sty or graphics.sty.}%
    \renewcommand\includegraphics[2][]{}%
  }%
  \providecommand\rotatebox[2]{#2}%
  \@ifundefined{ifGPcolor}{%
    \newif\ifGPcolor
    \GPcolortrue
  }{}%
  \@ifundefined{ifGPblacktext}{%
    \newif\ifGPblacktext
    \GPblacktexttrue
  }{}%
  % define a \g@addto@macro without @ in the name:
  \let\gplgaddtomacro\g@addto@macro
  % define empty templates for all commands taking text:
  \gdef\gplbacktexta{}%
  \gdef\gplfronttexta{}%
  \gdef\gplbacktextb{}%
  \gdef\gplfronttextb{}%
  \makeatother
  \ifGPblacktext
    % no textcolor at all
    \def\colorrgb#1{}%
    \def\colorgray#1{}%
  \else
    % gray or color?
    \ifGPcolor
      \def\colorrgb#1{\color[rgb]{#1}}%
      \def\colorgray#1{\color[gray]{#1}}%
      \expandafter\def\csname LTw\endcsname{\color{white}}%
      \expandafter\def\csname LTb\endcsname{\color{black}}%
      \expandafter\def\csname LTa\endcsname{\color{black}}%
      \expandafter\def\csname LT0\endcsname{\color[rgb]{1,0,0}}%
      \expandafter\def\csname LT1\endcsname{\color[rgb]{0,1,0}}%
      \expandafter\def\csname LT2\endcsname{\color[rgb]{0,0,1}}%
      \expandafter\def\csname LT3\endcsname{\color[rgb]{1,0,1}}%
      \expandafter\def\csname LT4\endcsname{\color[rgb]{0,1,1}}%
      \expandafter\def\csname LT5\endcsname{\color[rgb]{1,1,0}}%
      \expandafter\def\csname LT6\endcsname{\color[rgb]{0,0,0}}%
      \expandafter\def\csname LT7\endcsname{\color[rgb]{1,0.3,0}}%
      \expandafter\def\csname LT8\endcsname{\color[rgb]{0.5,0.5,0.5}}%
    \else
      % gray
      \def\colorrgb#1{\color{black}}%
      \def\colorgray#1{\color[gray]{#1}}%
      \expandafter\def\csname LTw\endcsname{\color{white}}%
      \expandafter\def\csname LTb\endcsname{\color{black}}%
      \expandafter\def\csname LTa\endcsname{\color{black}}%
      \expandafter\def\csname LT0\endcsname{\color{black}}%
      \expandafter\def\csname LT1\endcsname{\color{black}}%
      \expandafter\def\csname LT2\endcsname{\color{black}}%
      \expandafter\def\csname LT3\endcsname{\color{black}}%
      \expandafter\def\csname LT4\endcsname{\color{black}}%
      \expandafter\def\csname LT5\endcsname{\color{black}}%
      \expandafter\def\csname LT6\endcsname{\color{black}}%
      \expandafter\def\csname LT7\endcsname{\color{black}}%
      \expandafter\def\csname LT8\endcsname{\color{black}}%
    \fi
  \fi
  \begin{tabular}{c c}
  \setlength{\unitlength}{0.03675bp}%
  \begin{picture}(7200.00,5040.00)(500,0)%
    \gplgaddtomacro\gplbacktexta{%
      \csname LTb\endcsname%
      \put(1308,1587){\makebox(0,0)[r]{\strut{}-0.4}}%
      \put(1308,2242){\makebox(0,0)[r]{\strut{}-0.2}}%
      \put(1308,2898){\makebox(0,0)[r]{\strut{} 0}}%
      \put(1308,3553){\makebox(0,0)[r]{\strut{} 0.2}}%
      \put(1308,4208){\makebox(0,0)[r]{\strut{} 0.4}}%
      \put(1908,1039){\makebox(0,0){\strut{}-4}}%
      \put(2844,1039){\makebox(0,0){\strut{}-2}}%
      \put(3780,1039){\makebox(0,0){\strut{} 0}}%
      \put(4715,1039){\makebox(0,0){\strut{} 2}}%
      \put(5651,1039){\makebox(0,0){\strut{} 4}}%
      \put(538,2897){\rotatebox{-270}{\makebox(0,0){\strut{}$G_{\cal W}^l$~[TeV$^{-1}$]}}}%
      \put(3779,709){\makebox(0,0){\strut{}$G_{\cal W}^\phi$~[TeV$^{-1}$]}}%
    }%
    \gplgaddtomacro\gplfronttexta{%
    }%
    \gplgaddtomacro\gplbacktexta{%
    }%
    \gplgaddtomacro\gplfronttexta{%
    }%
    \gplgaddtomacro\gplbacktexta{%
    }%
    \gplgaddtomacro\gplfronttexta{%
    }%
    \gplgaddtomacro\gplbacktexta{%
    }%
    \gplgaddtomacro\gplfronttexta{%
    }%
    \gplgaddtomacro\gplbacktexta{%
    }%
    \gplgaddtomacro\gplfronttexta{%
    }%
    \gplbacktexta
    \put(0,0){\includegraphics[scale=0.735]{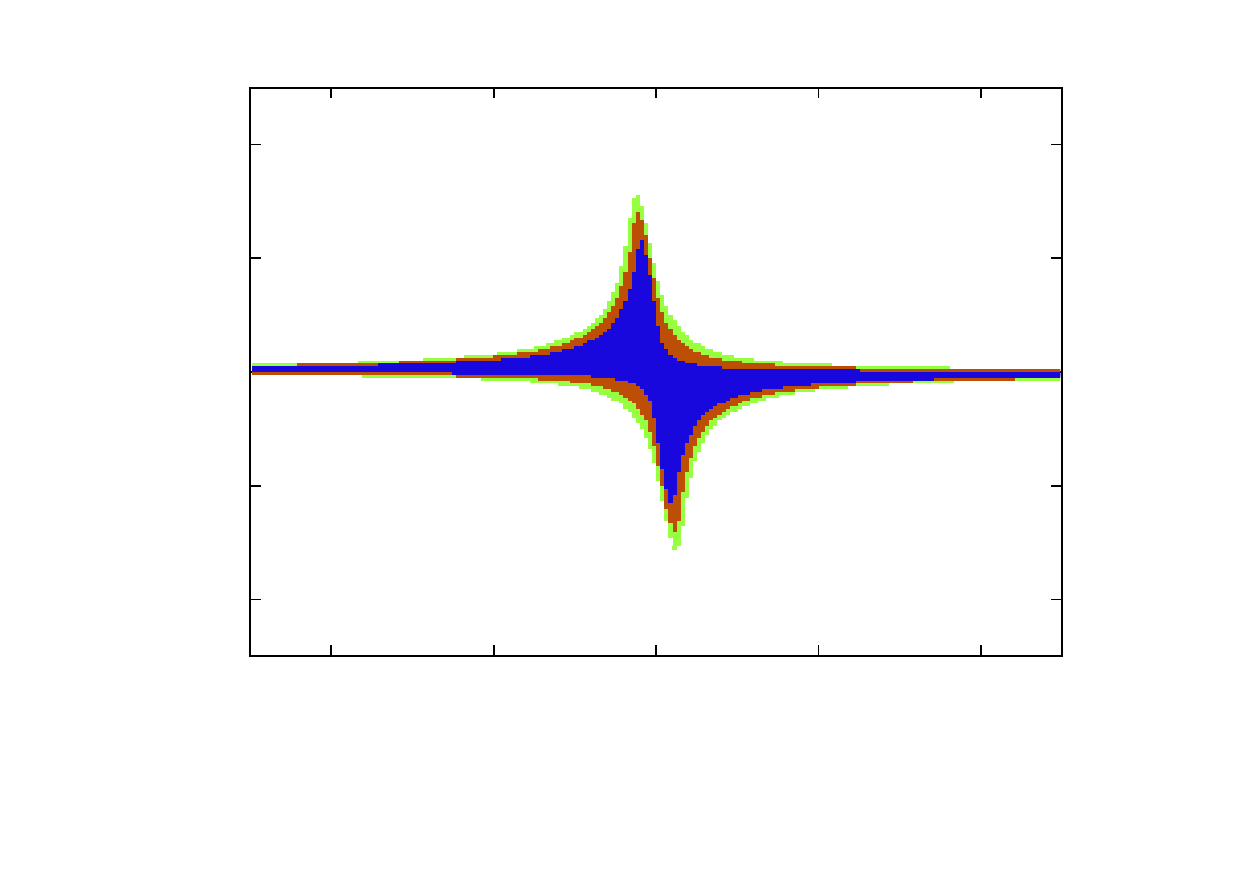}}%
    \gplfronttexta
  \end{picture}%
  &
  \setlength{\unitlength}{0.03675bp}%
  \begin{picture}(7200.00,5040.00)(1400,0)%
    \gplgaddtomacro\gplbacktextb{%
    }%
    \gplgaddtomacro\gplfronttextb{%
    }%
    \gplgaddtomacro\gplbacktextb{%
    }%
    \gplgaddtomacro\gplfronttextb{%
    }%
    \gplgaddtomacro\gplbacktextb{%
    }%
    \gplgaddtomacro\gplfronttextb{%
    }%
    \gplgaddtomacro\gplbacktextb{%
    }%
    \gplgaddtomacro\gplfronttextb{%
    }%
    \gplgaddtomacro\gplbacktextb{%
    }%
    \gplgaddtomacro\gplfronttextb{%
    }%
    \gplgaddtomacro\gplbacktextb{%
      \csname LTb\endcsname%
      \put(1308,1587){\makebox(0,0)[r]{\strut{}-0.4}}%
      \put(1308,2242){\makebox(0,0)[r]{\strut{}-0.2}}%
      \put(1308,2898){\makebox(0,0)[r]{\strut{} 0}}%
      \put(1308,3553){\makebox(0,0)[r]{\strut{} 0.2}}%
      \put(1308,4208){\makebox(0,0)[r]{\strut{} 0.4}}%
      \put(1908,1039){\makebox(0,0){\strut{}-4}}%
      \put(2844,1039){\makebox(0,0){\strut{}-2}}%
      \put(3780,1039){\makebox(0,0){\strut{} 0}}%
      \put(4715,1039){\makebox(0,0){\strut{} 2}}%
      \put(5651,1039){\makebox(0,0){\strut{} 4}}%
      \put(538,2897){\rotatebox{-270}{\makebox(0,0){\strut{}$G_{\cal W}^l$~[TeV$^{-1}$]}}}%
      \put(3779,709){\makebox(0,0){\strut{}$G_{\cal W}^q$~[TeV$^{-1}$]}}%
    }%
    \gplgaddtomacro\gplfronttextb{%
    }%
    \gplbacktextb
    \put(0,0){\includegraphics[scale=0.735]{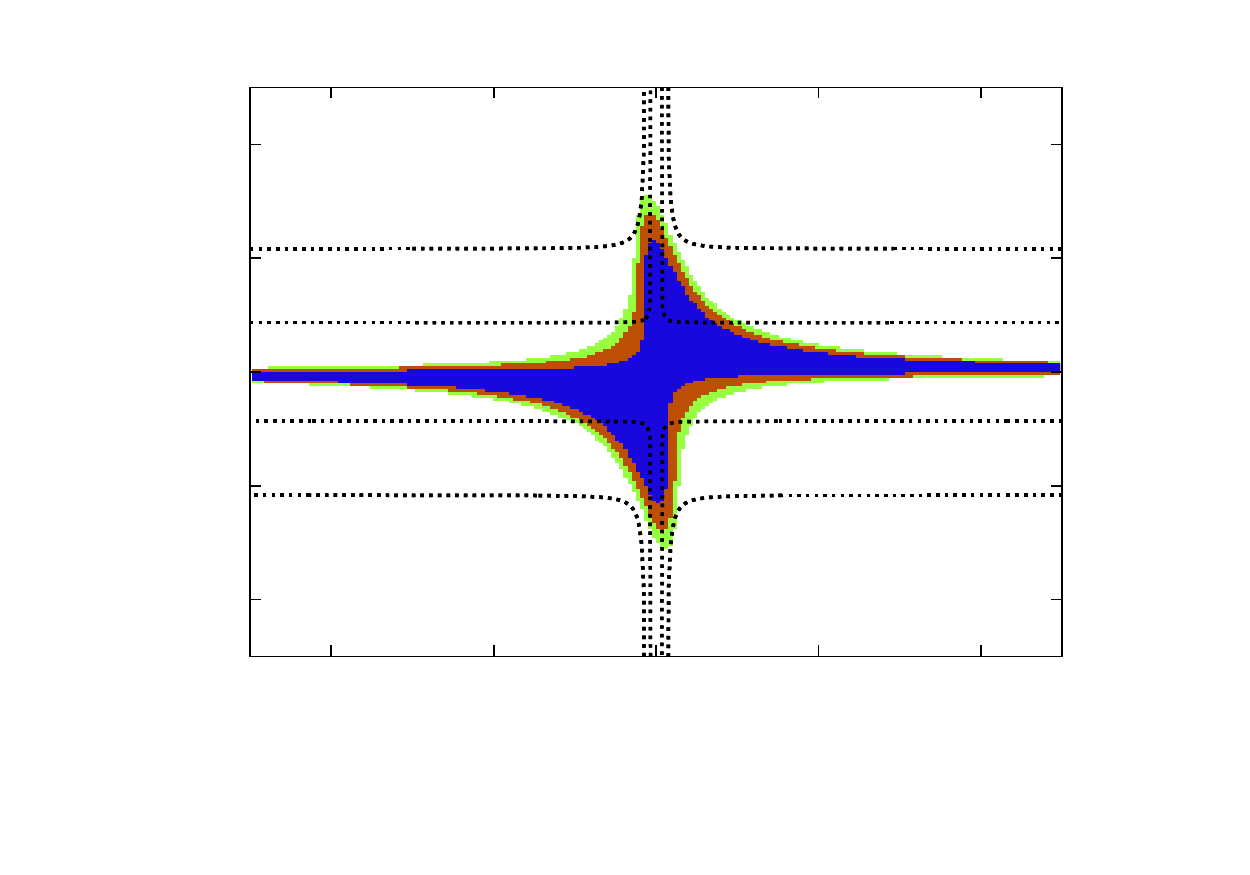}}%
    \gplfronttextb
  \end{picture}%  
  \end{tabular}
\endgroup

%% file: cucd_W0.tex
% GNUPLOT: LaTeX picture with Postscript
\begingroup
  \makeatletter
  \providecommand\color[2][]{%
    \GenericError{(gnuplot) \space\space\space\@spaces}{%
      Package color not loaded in conjunction with
      terminal option `colourtext'%
    }{See the gnuplot documentation for explanation.%
    }{Either use 'blacktext' in gnuplot or load the package
      color.sty in LaTeX.}%
    \renewcommand\color[2][]{}%
  }%
  \providecommand\includegraphics[2][]{%
    \GenericError{(gnuplot) \space\space\space\@spaces}{%
      Package graphicx or graphics not loaded%
    }{See the gnuplot documentation for explanation.%
    }{The gnuplot epslatex terminal needs graphicx.sty or graphics.sty.}%
    \renewcommand\includegraphics[2][]{}%
  }%
  \providecommand\rotatebox[2]{#2}%
  \@ifundefined{ifGPcolor}{%
    \newif\ifGPcolor
    \GPcolortrue
  }{}%
  \@ifundefined{ifGPblacktext}{%
    \newif\ifGPblacktext
    \GPblacktexttrue
  }{}%
  % define a \g@addto@macro without @ in the name:
  \let\gplgaddtomacro\g@addto@macro
  % define empty templates for all commands taking text:
  \gdef\gplbacktext{}%
  \gdef\gplfronttext{}%
  \makeatother
  \ifGPblacktext
    % no textcolor at all
    \def\colorrgb#1{}%
    \def\colorgray#1{}%
  \else
    % gray or color?
    \ifGPcolor
      \def\colorrgb#1{\color[rgb]{#1}}%
      \def\colorgray#1{\color[gray]{#1}}%
      \expandafter\def\csname LTw\endcsname{\color{white}}%
      \expandafter\def\csname LTb\endcsname{\color{black}}%
      \expandafter\def\csname LTa\endcsname{\color{black}}%
      \expandafter\def\csname LT0\endcsname{\color[rgb]{1,0,0}}%
      \expandafter\def\csname LT1\endcsname{\color[rgb]{0,1,0}}%
      \expandafter\def\csname LT2\endcsname{\color[rgb]{0,0,1}}%
      \expandafter\def\csname LT3\endcsname{\color[rgb]{1,0,1}}%
      \expandafter\def\csname LT4\endcsname{\color[rgb]{0,1,1}}%
      \expandafter\def\csname LT5\endcsname{\color[rgb]{1,1,0}}%
      \expandafter\def\csname LT6\endcsname{\color[rgb]{0,0,0}}%
      \expandafter\def\csname LT7\endcsname{\color[rgb]{1,0.3,0}}%
      \expandafter\def\csname LT8\endcsname{\color[rgb]{0.5,0.5,0.5}}%
    \else
      % gray
      \def\colorrgb#1{\color{black}}%
      \def\colorgray#1{\color[gray]{#1}}%
      \expandafter\def\csname LTw\endcsname{\color{white}}%
      \expandafter\def\csname LTb\endcsname{\color{black}}%
      \expandafter\def\csname LTa\endcsname{\color{black}}%
      \expandafter\def\csname LT0\endcsname{\color{black}}%
      \expandafter\def\csname LT1\endcsname{\color{black}}%
      \expandafter\def\csname LT2\endcsname{\color{black}}%
      \expandafter\def\csname LT3\endcsname{\color{black}}%
      \expandafter\def\csname LT4\endcsname{\color{black}}%
      \expandafter\def\csname LT5\endcsname{\color{black}}%
      \expandafter\def\csname LT6\endcsname{\color{black}}%
      \expandafter\def\csname LT7\endcsname{\color{black}}%
      \expandafter\def\csname LT8\endcsname{\color{black}}%
    \fi
  \fi
  \setlength{\unitlength}{0.0396bp}%
  \begin{picture}(6480.00,5443.20)%
\hspace{-0.5cm}
    \gplgaddtomacro\gplbacktext{%
      \csname LTb\endcsname%
      \put(376,2941){\rotatebox{-270}{\makebox(0,0){\strut{}\Large{$c_u$}}}}%
      \put(3580,154){\makebox(0,0){\strut{}\Large{$c_d$}}}%
      \put(1146,4986){\makebox(0,0)[l]{\strut{}\scriptsize{\textbf{3100 GeV}}}}%
      \put(1146,4754){\makebox(0,0)[l]{\strut{}\scriptsize{\textbf{3000 GeV}}}}%
      \put(1146,4508){\makebox(0,0)[l]{\strut{}\scriptsize{\textbf{2900 GeV}}}}%
      \put(1146,4258){\makebox(0,0)[l]{\strut{}\scriptsize{\textbf{2800 GeV}}}}%
      \put(1146,4001){\makebox(0,0)[l]{\strut{}\scriptsize{\textbf{2700 GeV}}}}%
      \put(1146,3754){\makebox(0,0)[l]{\strut{}\scriptsize{\textbf{2600 GeV}}}}%
      \put(1146,3517){\makebox(0,0)[l]{\strut{}\scriptsize{\textbf{2500 GeV}}}}%
      \put(1146,3271){\makebox(0,0)[l]{\strut{}\scriptsize{\textbf{2400 GeV}}}}%
      \put(1146,3013){\makebox(0,0)[l]{\strut{}\scriptsize{\textbf{2300 GeV}}}}%
      \put(1146,2772){\makebox(0,0)[l]{\strut{}\scriptsize{\textbf{2200 GeV}}}}%
      \put(1146,2539){\makebox(0,0)[l]{\strut{}\scriptsize{\textbf{2100 GeV}}}}%
      \put(1146,2282){\makebox(0,0)[l]{\strut{}\scriptsize{\textbf{2000 GeV}}}}%
      \put(1146,2032){\makebox(0,0)[l]{\strut{}\scriptsize{\textbf{1900 GeV}}}}%
      \put(2159,1543){\makebox(0,0)[l]{\strut{}\scriptsize{0.2}}}%
      \put(2735,2058){\makebox(0,0)[l]{\strut{}\scriptsize{0.3}}}%
      \put(3144,2423){\makebox(0,0)[l]{\strut{}\scriptsize{0.4}}}%
      \put(3461,2706){\makebox(0,0)[l]{\strut{}\scriptsize{0.5}}}%
      \put(3720,2938){\makebox(0,0)[l]{\strut{}\scriptsize{0.6}}}%
      \put(3939,3134){\makebox(0,0)[l]{\strut{}\scriptsize{0.7}}}%
      \put(4128,3303){\makebox(0,0)[l]{\strut{}\scriptsize{0.8}}}%
      \put(4296,3453){\makebox(0,0)[l]{\strut{}\scriptsize{0.9}}}%
      \put(4445,3587){\makebox(0,0)[l]{\strut{}\scriptsize{1.0}}}%
      \put(5430,4467){\makebox(0,0)[l]{\strut{}\scriptsize{2.0}}}%
      \put(5863,4907){\makebox(0,0)[l]{\strut{}\scriptsize{3.0}}}%
    }%
    \gplgaddtomacro\gplfronttext{%
      \csname LTb\endcsname%
      \put(2088,1225){\makebox(0,0)[r]{\strut{}$\mathcal{W}$\hspace{0.3cm}}}%
      \csname LTb\endcsname%
      \put(2088,983){\makebox(0,0)[r]{\strut{}$\tilde{g}=g$}}%
      \csname LTb\endcsname%
      \put(946,704){\makebox(0,0)[r]{\strut{}$10^{-4}$}}%
      \put(946,2166){\makebox(0,0)[r]{\strut{}$10^{-3}$}}%
      \put(946,3628){\makebox(0,0)[r]{\strut{}$10^{-2}$}}%
      \put(946,5089){\makebox(0,0)[r]{\strut{}$10^{-1}$}}%
      \put(1078,484){\makebox(0,0){\strut{}$10^{-4}$}}%
      \put(2713,484){\makebox(0,0){\strut{}$10^{-3}$}}%
      \put(4348,484){\makebox(0,0){\strut{}$10^{-2}$}}%
      \put(5984,484){\makebox(0,0){\strut{}$10^{-1}$}}%
    }%
    \gplbacktext
    \put(0,0){\includegraphics[scale=0.792]{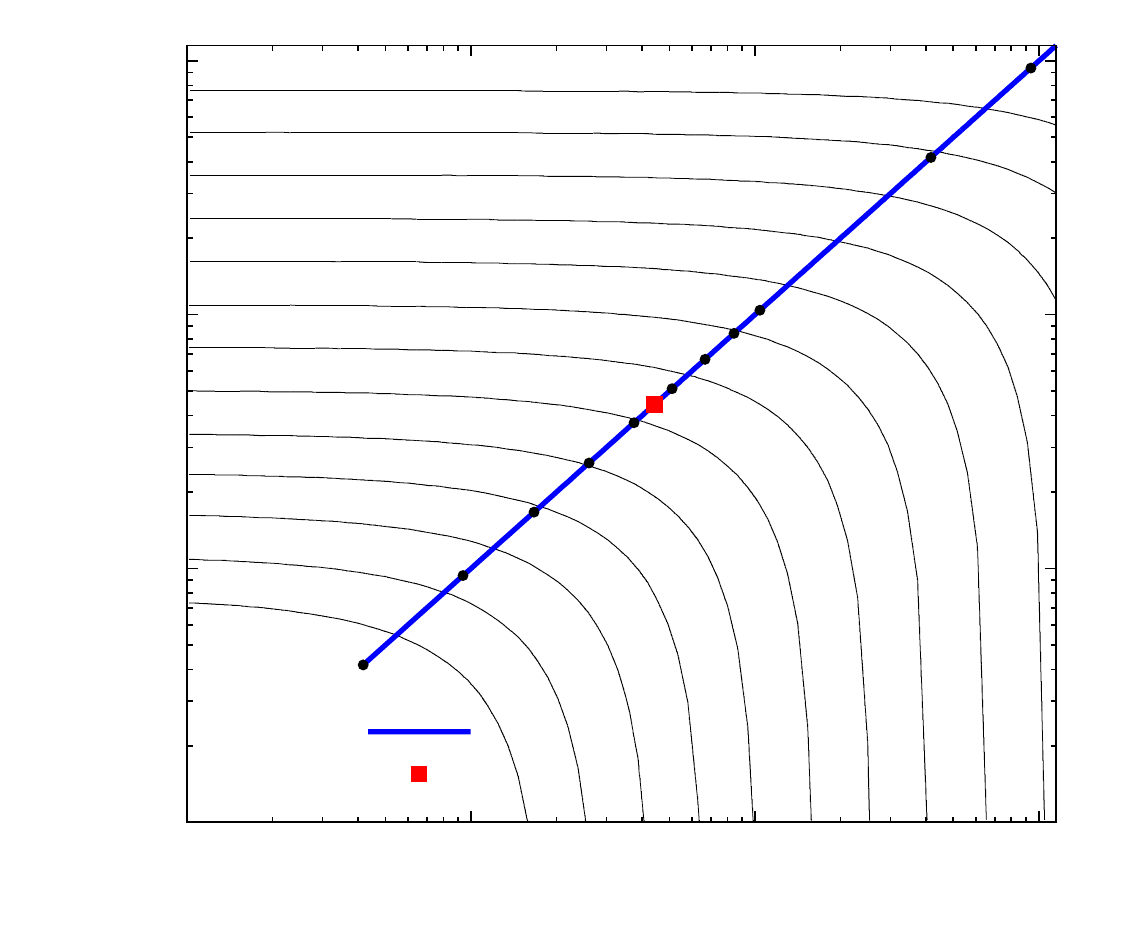}}%
    \gplfronttext
  \end{picture}%
\endgroup

%% file: limtrp_gH0.tex
% GNUPLOT: LaTeX picture with Postscript
\begingroup
  \makeatletter
  \providecommand\color[2][]{%
    \GenericError{(gnuplot) \space\space\space\@spaces}{%
      Package color not loaded in conjunction with
      terminal option `colourtext'%
    }{See the gnuplot documentation for explanation.%
    }{Either use 'blacktext' in gnuplot or load the package
      color.sty in LaTeX.}%
    \renewcommand\color[2][]{}%
  }%
  \providecommand\includegraphics[2][]{%
    \GenericError{(gnuplot) \space\space\space\@spaces}{%
      Package graphicx or graphics not loaded%
    }{See the gnuplot documentation for explanation.%
    }{The gnuplot epslatex terminal needs graphicx.sty or graphics.sty.}%
    \renewcommand\includegraphics[2][]{}%
  }%
  \providecommand\rotatebox[2]{#2}%
  \@ifundefined{ifGPcolor}{%
    \newif\ifGPcolor
    \GPcolortrue
  }{}%
  \@ifundefined{ifGPblacktext}{%
    \newif\ifGPblacktext
    \GPblacktexttrue
  }{}%
  % define a \g@addto@macro without @ in the name:
  \let\gplgaddtomacro\g@addto@macro
  % define empty templates for all commands taking text:
  \gdef\gplbacktext{}%
  \gdef\gplfronttext{}%
  \makeatother
  \ifGPblacktext
    % no textcolor at all
    \def\colorrgb#1{}%
    \def\colorgray#1{}%
  \else
    % gray or color?
    \ifGPcolor
      \def\colorrgb#1{\color[rgb]{#1}}%
      \def\colorgray#1{\color[gray]{#1}}%
      \expandafter\def\csname LTw\endcsname{\color{white}}%
      \expandafter\def\csname LTb\endcsname{\color{black}}%
      \expandafter\def\csname LTa\endcsname{\color{black}}%
      \expandafter\def\csname LT0\endcsname{\color[rgb]{1,0,0}}%
      \expandafter\def\csname LT1\endcsname{\color[rgb]{0,1,0}}%
      \expandafter\def\csname LT2\endcsname{\color[rgb]{0,0,1}}%
      \expandafter\def\csname LT3\endcsname{\color[rgb]{1,0,1}}%
      \expandafter\def\csname LT4\endcsname{\color[rgb]{0,1,1}}%
      \expandafter\def\csname LT5\endcsname{\color[rgb]{1,1,0}}%
      \expandafter\def\csname LT6\endcsname{\color[rgb]{0,0,0}}%
      \expandafter\def\csname LT7\endcsname{\color[rgb]{1,0.3,0}}%
      \expandafter\def\csname LT8\endcsname{\color[rgb]{0.5,0.5,0.5}}%
    \else
      % gray
      \def\colorrgb#1{\color{black}}%
      \def\colorgray#1{\color[gray]{#1}}%
      \expandafter\def\csname LTw\endcsname{\color{white}}%
      \expandafter\def\csname LTb\endcsname{\color{black}}%
      \expandafter\def\csname LTa\endcsname{\color{black}}%
      \expandafter\def\csname LT0\endcsname{\color{black}}%
      \expandafter\def\csname LT1\endcsname{\color{black}}%
      \expandafter\def\csname LT2\endcsname{\color{black}}%
      \expandafter\def\csname LT3\endcsname{\color{black}}%
      \expandafter\def\csname LT4\endcsname{\color{black}}%
      \expandafter\def\csname LT5\endcsname{\color{black}}%
      \expandafter\def\csname LT6\endcsname{\color{black}}%
      \expandafter\def\csname LT7\endcsname{\color{black}}%
      \expandafter\def\csname LT8\endcsname{\color{black}}%
    \fi
  \fi
  \setlength{\unitlength}{0.0375bp}%
  \begin{picture}(7800.00,5040.00)%
\hspace{-0.95cm}
    \gplgaddtomacro\gplbacktext{%
      \csname LTb\endcsname%
      \put(500,3260){\rotatebox{-270}{\makebox(0,0){\strut{}$\tilde{g}$}}}%
      \put(3940,154){\makebox(0,0){\strut{}$M_{\mathcal{W}}$ [GeV]}}%
    }%
    \gplgaddtomacro\gplfronttext{%
      \csname LTb\endcsname%
      \put(1651,1350){\rotatebox{6}{\makebox(0,0)[l]{\strut{}\small{95$\%$ C.L. EW limits}}}}%
      \csname LTb\endcsname%
      \put(5089,4123){\makebox(0,0)[r]{\strut{}\small{Combined~~~}}}%
      \csname LTb\endcsname%
      \put(5089,3837){\makebox(0,0)[r]{\strut{}\small{Lepton+MET (CMS, 7 TeV, 5 fb$^\tinymath{-1}$)~~}}}%
      \csname LTb\endcsname%
      \put(5089,3551){\makebox(0,0)[r]{\strut{}\small{Dilepton (CMS, 7 TeV, 5 fb$^\tinymath{-1}$)~~}}}%
      \csname LTb\endcsname%
      \put(946,704){\makebox(0,0)[r]{\strut{} 0}}%
      \put(946,1213){\makebox(0,0)[r]{\strut{} 0.2}}%
      \put(946,1722){\makebox(0,0)[r]{\strut{} 0.4}}%
      \put(946,2231){\makebox(0,0)[r]{\strut{} 0.6}}%
      \put(946,2740){\makebox(0,0)[r]{\strut{} 0.8}}%
      \put(946,3248){\makebox(0,0)[r]{\strut{} 1}}%
      \put(946,3757){\makebox(0,0)[r]{\strut{} 1.2}}%
      \put(946,4266){\makebox(0,0)[r]{\strut{} 1.4}}%
      \put(946,4775){\makebox(0,0)[r]{\strut{} 1.6}}%
      \put(1078,484){\makebox(0,0){\strut{} 1000}}%
      \put(2509,484){\makebox(0,0){\strut{} 1500}}%
      \put(3941,484){\makebox(0,0){\strut{} 2000}}%
      \put(5372,484){\makebox(0,0){\strut{} 2500}}%
      \put(6803,484){\makebox(0,0){\strut{} 3000}}%
    }%
    \gplbacktext
    \put(0,0){\includegraphics[scale=0.75]{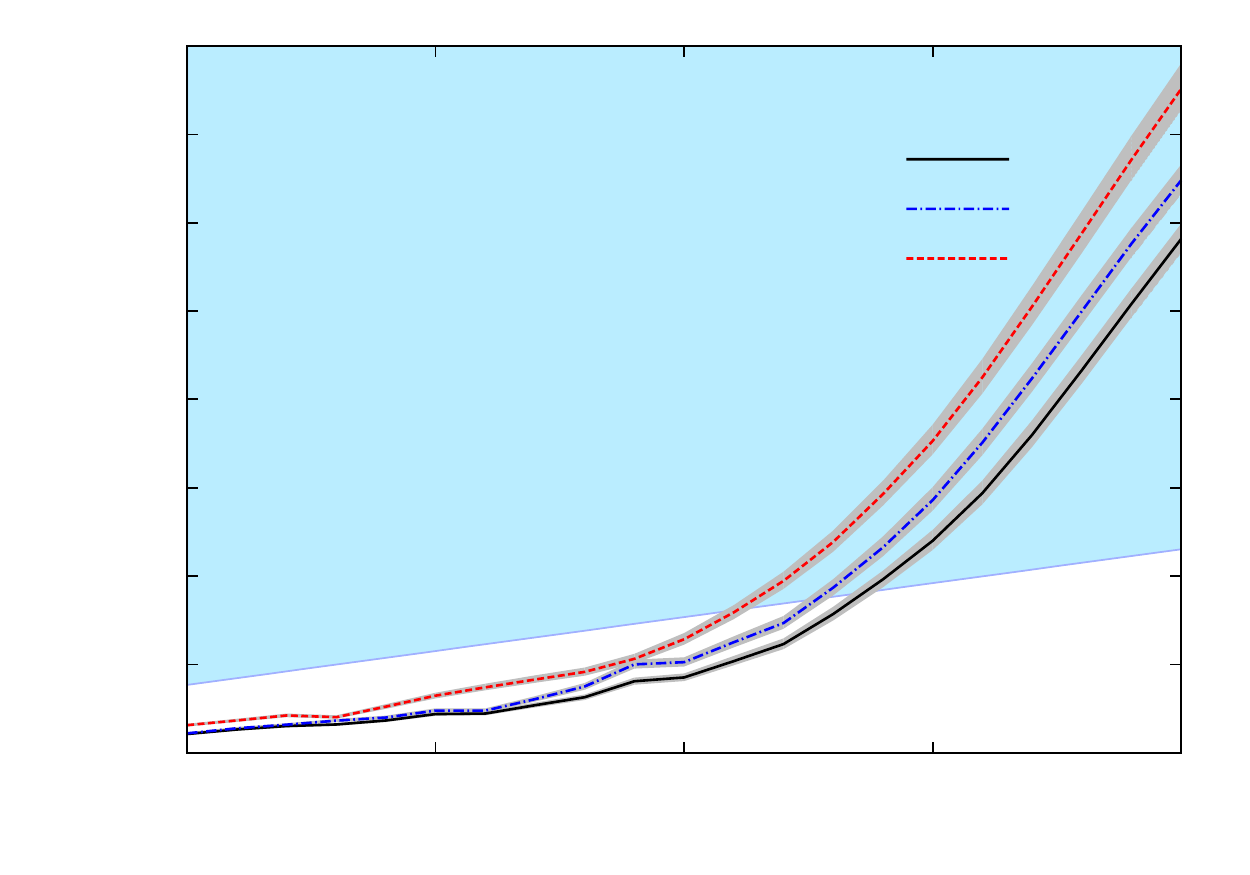}}%
    \gplfronttext
  \end{picture}%
\endgroup